\documentclass[10pt, conference, letterpaper]{IEEEtran}
\usepackage{graphicx} 

\title{See Where You Read with Eye Gaze Tracking and Large Language Model}

\usepackage{eucal}
\usepackage{bm}
\usepackage{makecell}
\usepackage{algorithm}
\usepackage{lineno}
\usepackage{algpseudocode}
\usepackage{amsmath}
\usepackage{subfigure}
\usepackage{booktabs}
\usepackage{multirow}
\usepackage{enumitem}
\usepackage{flowchart}
\usepackage{float}
\usepackage{times}
\usepackage{url}
\usepackage{amsfonts}

\usepackage{amssymb}
\usepackage{xspace}
\usepackage{graphicx}
\usepackage{tabu}
\usepackage{diagbox}
\usepackage{caption}
\usepackage{subcaption}
\usepackage{xcolor}

\usepackage[T1]{fontenc}
\usepackage{helvet}
\pdfmapfile{+helvetica.map}

\newcommand{\ours}{\textsc{$RT^2H$}\xspace}

\author{
    \IEEEauthorblockN{Sikai Yang}
    \IEEEauthorblockA{
        University of California, Merced \\
        syang126@ucmerced.edu}
    \and
    \IEEEauthorblockN{Gang Yan}
    \IEEEauthorblockA{
        University of California, Merced \\
        gyan5@ucmerced.edu}
    \and
    \IEEEauthorblockN{Wan Du}
    \IEEEauthorblockA{
        University of California, Merced \\
        wdu3@ucmerced.edu}
}

\date{\today}

\begin{document}
    \maketitle
    \pagestyle{plain}
    \section*{Abstract}

Losing track of reading progress during line switching can be frustrating.
Eye gaze tracking technology offers a potential solution by highlighting read paragraphs, aiding users in avoiding wrong line switches. 
However, the gap between gaze tracking accuracy (2-3 cm) and text line spacing (3-5 mm) makes direct application impractical. 
Existing methods leverage the linear reading pattern but fail during jump reading.
This paper presents a reading tracking and highlighting system that supports both linear and jump reading. 
Based on experimental insights from the gaze nature study of 16 users, two gaze error models are designed to enable both jump reading detection and relocation.
The system further leverages the large language model's contextual perception capability in aiding reading tracking.
A reading tracking domain-specific line-gaze alignment opportunity is also exploited to enable dynamic and frequent calibration of the gaze results.
Controlled experiments demonstrate reliable linear reading tracking, as well as 84\% accuracy in tracking jump reading. 
Furthermore, real field tests with 18 volunteers demonstrated the system’s effectiveness in tracking and highlighting read paragraphs, improving reading efficiency, and enhancing user experience.

    \section{introduction}

Losing track of reading progress can be annoying.
This often happens when a reader tries to move to a new line after finishing the previous one but accidentally switches to the wrong line, requiring them to double-check the sentences and find the correct spot.

Advancements in eye gaze tracking technology \cite{majaranta2014eye,ASgaze,Balim_2023_CVPR,DVgaze} have provided new opportunities to assist with reading and address this issue.
By tracking reading progress through gaze and highlighting read paragraphs, users receive visual assistance that prevents them from switching to the wrong line.
This technology also facilitates other applications  \cite{palinko2010estimating,ren2011affective}, such as person tracking \cite{nalaie2024learning}.
In this paper, we discuss the prospect of enabling practical and robust reading tracking application using existing gaze tracking technology (e.g., Tobii Eye Tracking \cite{tobii}).

However, there is a significant gap between gaze tracking accuracy and the level required for reliable reading tracking.
Typical gaze tracking accuracy ranges from 2 to 3 centimeters \cite{majaranta2014eye, holmqvist2011eye, duchowski2007eye}, while the spacing between lines in reading materials is typically 3 to 5 millimeters.
This discrepancy makes it impractical to directly apply gaze tracking results for tracking reading progress.
Unfortunately, this gap is challenging to reduce, since gaze tracking accuracy is limited by the discontinuous nature of human gaze \cite{rayner1998eye, just1980theory, rayner2012psychology}.
Humans do not read continuously; instead, they read in saccades and fixations, with their eyes briefly scanning segments of a sentence without focusing on every word or letter.
Typical angular difference between fixations and saccades ranges from 2 to 3 degrees, which could cover 7 to 9 letters \cite{rayner200935th}, thereby limiting the accuracy of gaze tracking during reading. 
Given a distance of 50 cm from the screen to the user, the uncertainty can be approximately 2 cm.

\begin{figure}[t]
    \hfill
    \begin{minipage}[t]{0.48\linewidth}
        \centering
        \includegraphics[width=\textwidth]{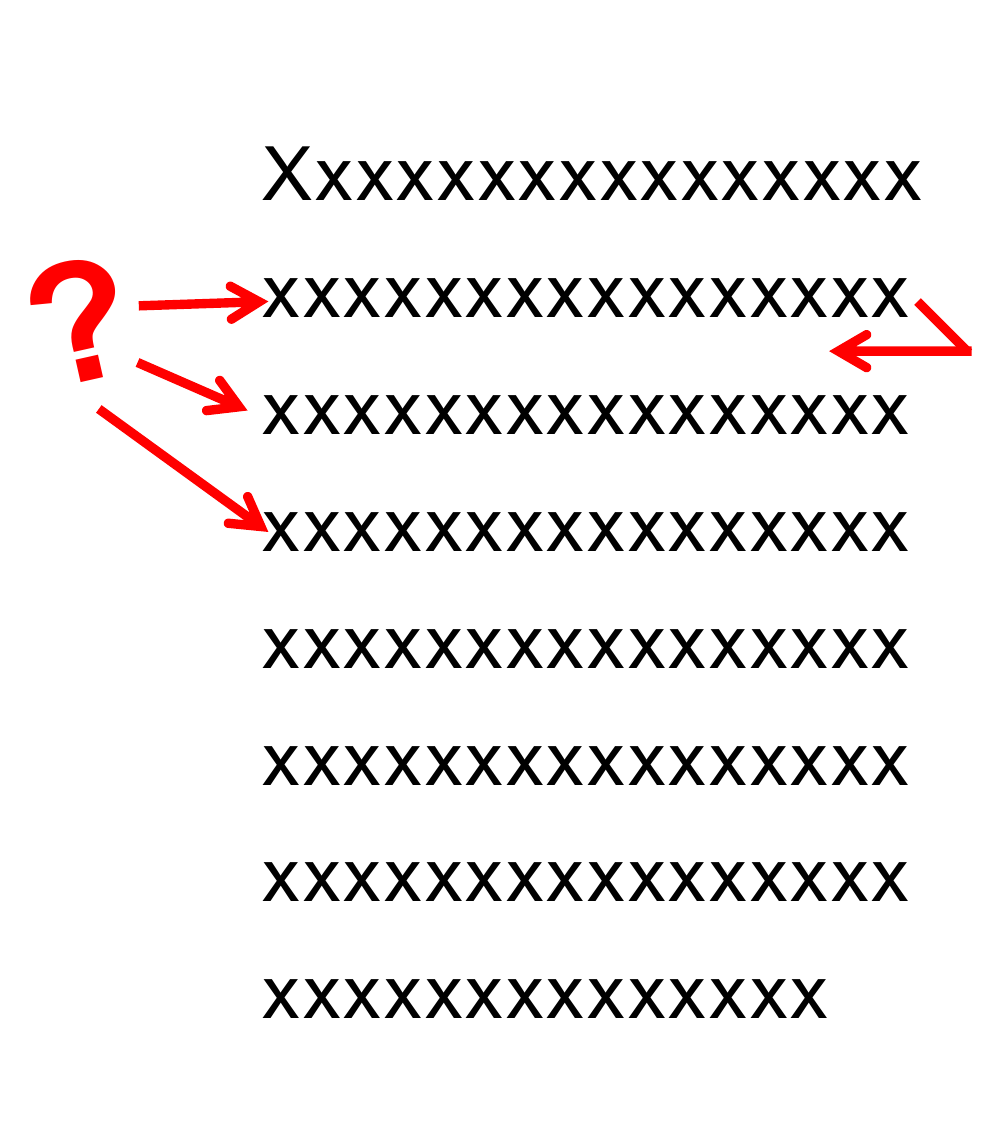}
         \vspace{-6mm}
        \caption{Losing track of reading progress}
        \label{fig_intro_1}
    \end{minipage}
    \hfill
    \begin{minipage}[t]{0.45\linewidth}
        \centering
        \includegraphics[width=\textwidth]{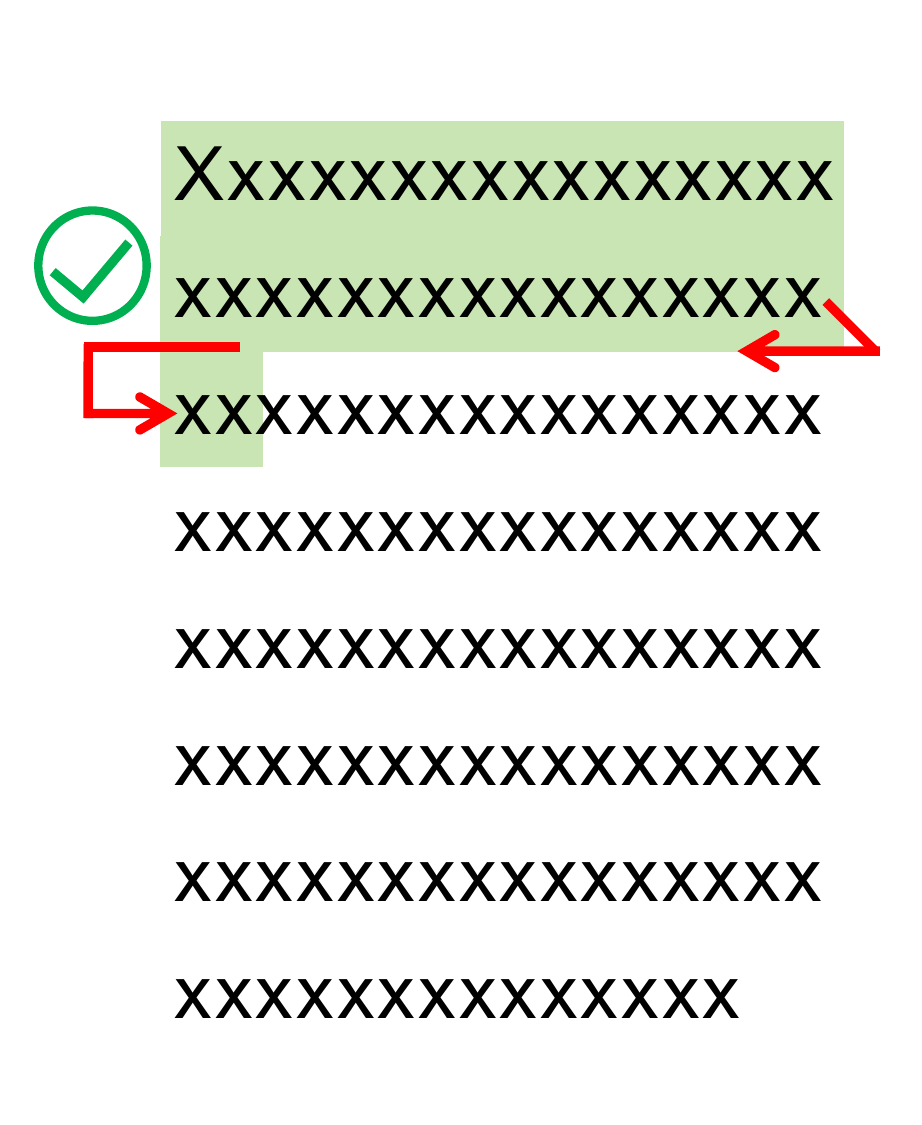}
         \vspace{-6mm}
        \caption{Reading assistance via highlighting}
        \label{fig_intro_2}
    \end{minipage}
    \vspace{-3mm}
    \hfill
    \hfill
\end{figure}

The remaining practical solution is to bridge the gap using constraints specific to reading tracking.
Existing reading tracking solutions \cite{bottos2019novel, bottos2019tracking, bottos2020tracking} leverage the linear reading pattern, where individuals read line by line.
This assumption provides strong constraints for reading tracking, effectively covering the gap between the fine line spacing and the relatively large gaze error.

However, a practical reading tracking system should also support jump reading, which consists of review and preview.
Users may review previously read paragraphs to verify certain concepts or information \cite{hornbaek2003reading}.
Additionally, they may also preview upcoming paragraphs if they feel fatigued by the current section and wish to jump ahead.
 In both scenarios, where users do not follow the line-by-line reading pattern, linear reading tracking methods are likely to fail.

In this work, we present the first reading tracking system \ours (\textbf{R}eading \textbf{T}racking and \textbf{R}eal-\textbf{T}ime \textbf{H}ighlighting) that supports jump reading, in addition to linear reading.
For linear reading tracking, we directly use the horizontal gaze position to determine the reading progress along the current line.
In addition to linear reading tracking, we detect jump reading by monitoring whether the gaze remains outside the normal error range of the current line for an extended period.
We collected gaze data from 16 users to determine the gaze error range and its distribution on the screen.
Using this distribution as reference, we dynamically adjust the estimated gaze error range to detect jump reading.
Once jump reading is detected, the system executes a relocation procedure to redirect the reading tracking progress to the new starting point and resumes linear reading tracking.
In designing the system, we address three major challenges with corresponding solutions:

\textbf{\textit{Jump Reading Relocation}}:
Relocating jump reading can be difficult, as gaze tracking error largely exceeds the line spacing, making it difficult to pinpoint the exact new line the user is reading.
To provide guidance for the relocation, we observe an opportunity of using punctuation marks as relocation anchors, since each sentence divided by punctuation marks represents a complete linguistic unit. 
Users are more likely to follow punctuation marks for efficient review or preview.
Therefore, when jump reading is detected, we track the user's gaze trajectory and search for punctuation marks within the error range of this trajectory to identify potential relocation candidates.
If there is only one candidate found within the error range of the trajectory, we use it as the new starting point. 
When multiple candidates are identified, the situation becomes more complex, requiring a suitable mechanism for candidate selection.

\textbf{\textit{LLM Assisted Candidate Election}}:
When multiple candidates are identified, it is essential to determine the best option using both gaze tracking results and contextual information.
Each punctuation mark candidate is followed by a sentence.
We first calculate the match ratio between the candidate sentences' locations and the gaze trajectory.
Specifically, we propose a finer probabilistic gaze error distribution model for the calculation, which is also developed from the gaze data collected from the 16 users.
This match ratio will serve as a major metric for evaluating the candidates.

Secondly, since we are dealing with reading materials, contextual information can also provide guidance.
Recent advancements in large language models (LLMs) offer new possibilities for enhancing reading tracking with language perception capabilities.
However, collecting a large ground truth dataset without disrupting users' natural reading patterns is challenging, making it impractical to fine-tune the LLM with domain-specific data \cite{huang2023towards}.
On the other hand, the original language perception capability of the LLM is adequate for understanding the reading material and assisting in the candidate election.
As a result, we employ e prompt engineering techniques to effectively utilize the LLM \cite{siledar2024one,hu2024improving,kan2023prompt,xu2024prompting, dxz_llm_1, dxz_llm_2}.
Based on the most recent linear reading history and the sentences following the punctuation mark candidates, we query the LLM to identify the most probable next reading candidate based on contextual relevance and logic. 
The candidate selected by the LLM receives an evaluation bonus during the election.
Putting together the trajectory-based evaluation and the LLM evaluation, the final punctuation mark elected from all candidates will serve as the new start point to resume linear reading tracking.

\textbf{\textit{Gaze Drifting and Dynamic Calibration}}:
Reading can be a time-consuming task, during which gaze tracking calibration may gradually degrade due to factors such as user repositioning.
This causes the gaze tracking result to drift away from its original accuracy, necessitating periodic re-calibration to correct gaze tracking results.
Classical gaze tracking calibration methods require active user attention, which will interrupt the current reading activity and demand the user's participation.
Fortunately, we identified a valuable opportunity for calibration uniquely suited to the reading tracking domain
Similar to the idea of saliency-based calibration \cite{yang2021vgaze, yang2022continuous}, we leverage the assumption that during linear reading, the user is actively focusing on the current line.
The alignment between the gaze and the line location provides valuable dynamic calibration resources during reading without specific user participation.
Specifically, we focus on calibrating the Y-axis (vertical) of gaze estimations, as it is the bottleneck of reading tracking.
For each completed line, we calculate the average Y-axis gaze locations and compare them with the vertical location of the finished line.
With multiple gaze-line pairs, we perform linear regression to fit the gaze estimations to the lines' vertical location.
This approach enables continuous calibration during linear reading and also after jump reading re-locations, ensuring consistent tracking accuracy.

We first conduct controlled experiments to statistically evaluate our system, in which we collect gaze ground truth via looking at the cursor and simulate reading.
Results show that \ours achieves 84\% accuracy in jump reading tracking.
However, cursor-based ground truth collection disrupts natural reading patterns, and simulated reading does not fully represent real-world scenarios.
To address these limitations, we further conduct a field test to put the system in real use case, as well as examining the prospect of reading assistance via text highlighting.

We invited 18 volunteers to participate in the test.
We use the reading tracking result to highlight the read paragraphs, thereby helping users visually track their progress and avoid switching to incorrect lines.
We evaluate the reading efficiency by clocking the time spent reading and comprehending specific paragraphs.
Results show \ours could reduce the required time by 13.5\% compared to the baseline methods. 
Additionally, we assessed user experience through questionnaires, which revealed that our jump reading tracking design is beneficial, with users expressing strong interest in having our system available commercially.


In summary, we mainly make the following contributions in designing the reading tracking and highlighting system:
\begin{itemize}
    \item To build a practical reading tracking system, we design a novel relocation mechanism to support jump reading.
    \item We introduce the usage of large language models in assisting the candidate election in jump reading relocation.
    \item We exploit the line-gaze alignment opportunities for dynamic calibration of the gaze tracking results.
    \item We implement the system, and conduct both controlled experiments and field tests for holistic evaluation.
\end{itemize}
    \section{Design}
In this section we introduce the design details of our reading tracking system \ours.
Section \ref{sec_desgin_linear} introduces the basic methods used for linear tracking.
Section \ref{sec_desgin_detection} and \ref{sec_desgin_relocation} introduce the techniques designed to handle the complex situations of jump reading.
Lastly, Section \ref{sec_desgin_calibration} demonstrates the possibility of exploiting a reading tracking domain-specific opportunity, i.e., the gaze-line alignment during linear reading, to enhance gaze tracking accuracy and stability.

\subsection{Linear Reading Tracking} \label{sec_desgin_linear}
In this work we mainly focus on handling jump reading, and design simple but effective linear reading tracking methods.
Text lines are extremely slender objects placed horizontally.
The millimeter-grade line height poses a great challenge for centimeter-grade gaze tracking.
Fortunately, the linear reading assumption greatly mitigates this challenge on the vertical aspect.
It represents the regular reading pattern, where the user reads from left to right within each line, and line after line.

\subsubsection{Horizontal Tracking}
For linear reading within each line, we directly use the horizontal gaze location as the reading progress of the line, since the horizontal scale of the text line largely exceeds the gaze tracking error.

\subsubsection{Vertical Tracking and Z-cut} \label{sec_z_cut}
Humans read from left to right.
If the gaze moves from right to left for a long distance, it highly likely indicates the user has finished reading a line and is moving to read the next line.
Therefore, for the line switching, we rely on the detection of `Z-cut', namely, the user reaches the right border and then returns to the left border to start reading a new line.
Due to the fact that humans read in saccades and fixations, the user's gaze may not exactly pass the border.
Therefore, we set a closer threshold for both left and right border checkpoints (e.g., 20\% of line width).

\begin{figure}[ht]
    \hfill
    \begin{minipage}[t]{0.9\linewidth}
        \centering
        \includegraphics[width=\textwidth]{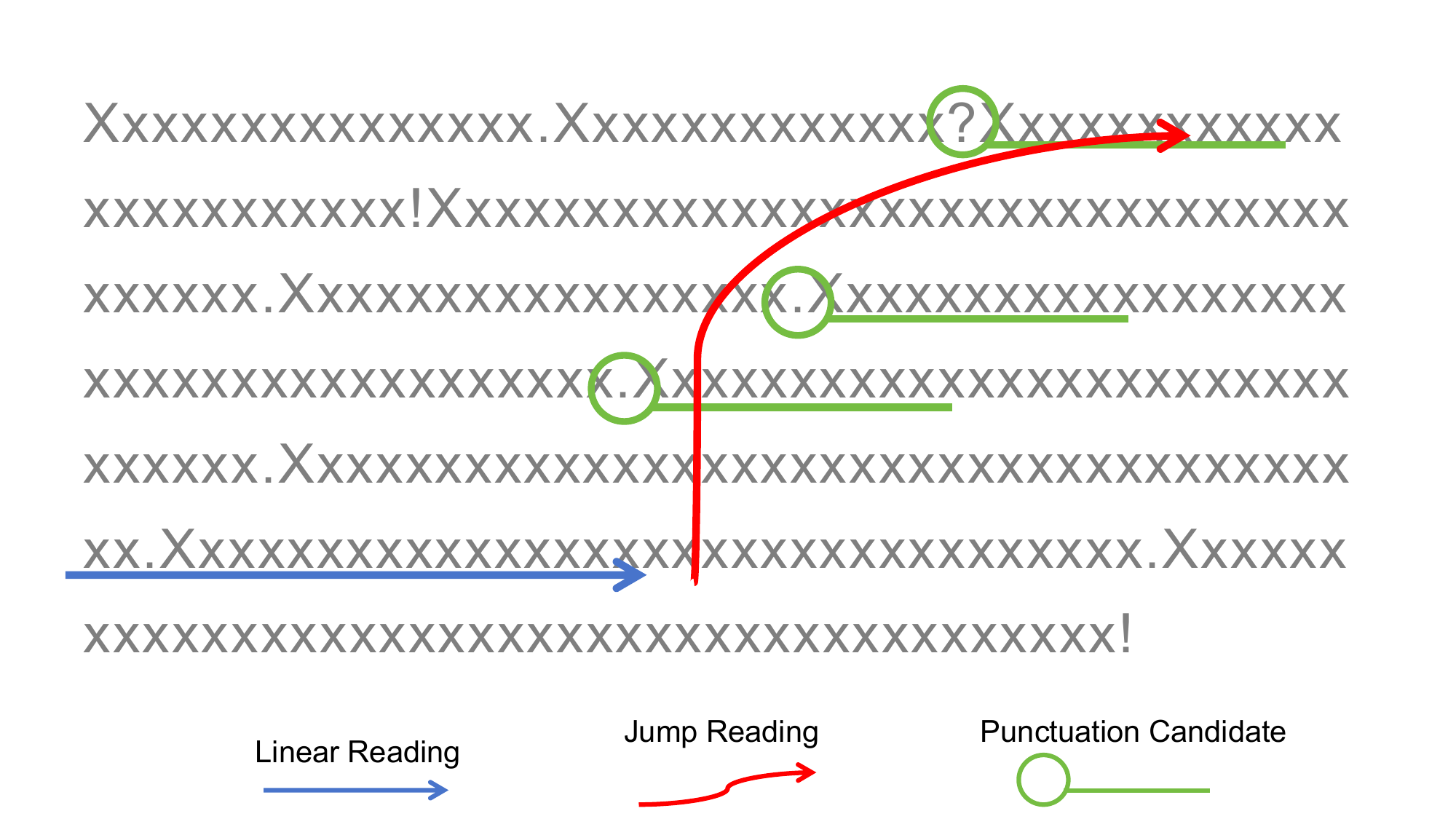}
         \vspace{-6mm}
        \caption{Reading tracking illustration}
        \label{fig_jump}
    \end{minipage}
    \hfill
    \hfill
\end{figure}

\subsection{Jump Reading Detection} \label{sec_desgin_detection}
Jump reading is much more challenging than linear reading as it breaks the line-by-line reading assumption.
It includes review, where the user returns to text that has been read, and preview, where the user skips ahead to read other information, as illustrated in Figure \ref{fig_jump}.
In handling jump reading, as the first step, we detect jump reading via monitoring whether the gaze escapes the current line for too long.
From the temporal aspect, we set a time threshold of 2.5 seconds of accumulative active gazing to determine if jump reading is initiated.
Specifically, if the gaze is not detected, it will not be accumulated into the threshold timer, as it probably means the user is looking away from the screen.

From the spatial aspect, we monitor if the gaze escapes a certain range from the line that is currently being read.
However, setting this range is not straightforward, as it is expected to suit different users' gaze patterns.
To that end, we rely on a real user study to generalize a gaze error range model.

\begin{figure}[ht]
    \hfill
    \begin{minipage}[t]{0.99\linewidth}
        \centering
        \includegraphics[width=\textwidth]{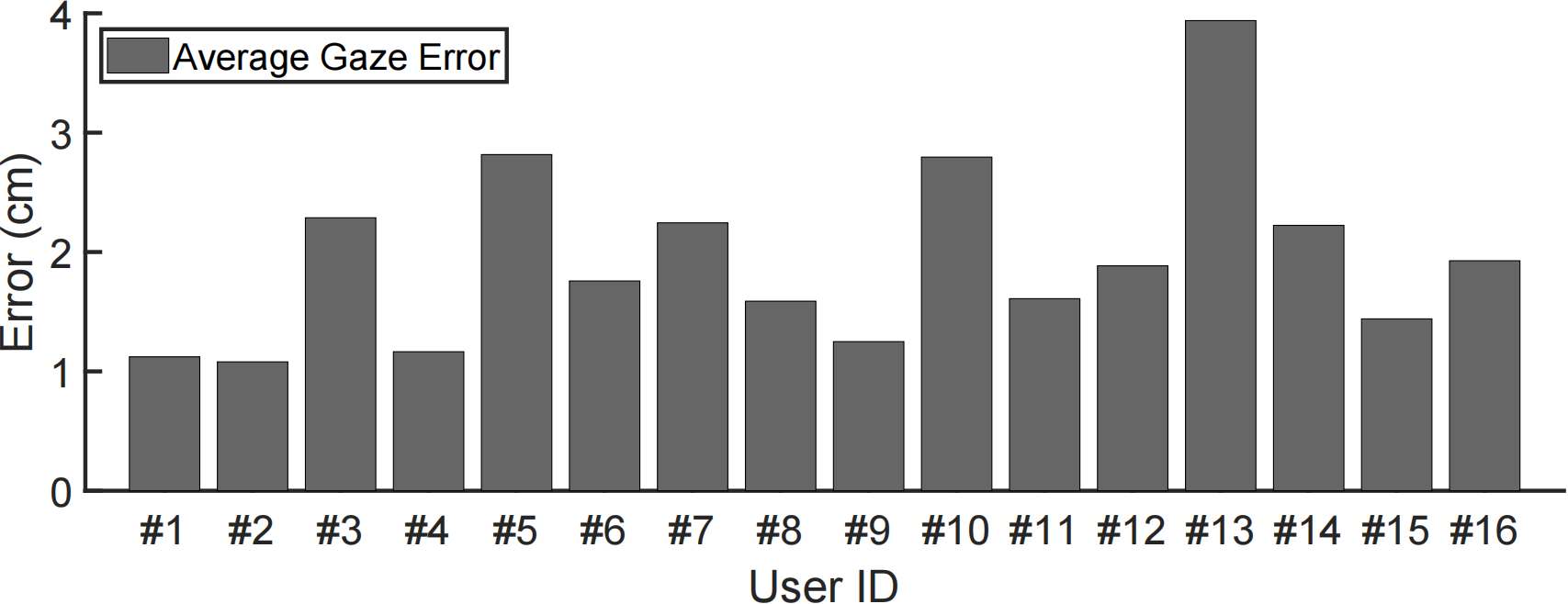}
         \vspace{-6mm}
        \caption{Gaze error among participants}
        \label{fig_users}
    \end{minipage}
    \hfill
\end{figure}

\begin{figure}[ht]
    \hfill
    \begin{minipage}[t]{0.98\linewidth}
        \centering
        \includegraphics[width=\textwidth]{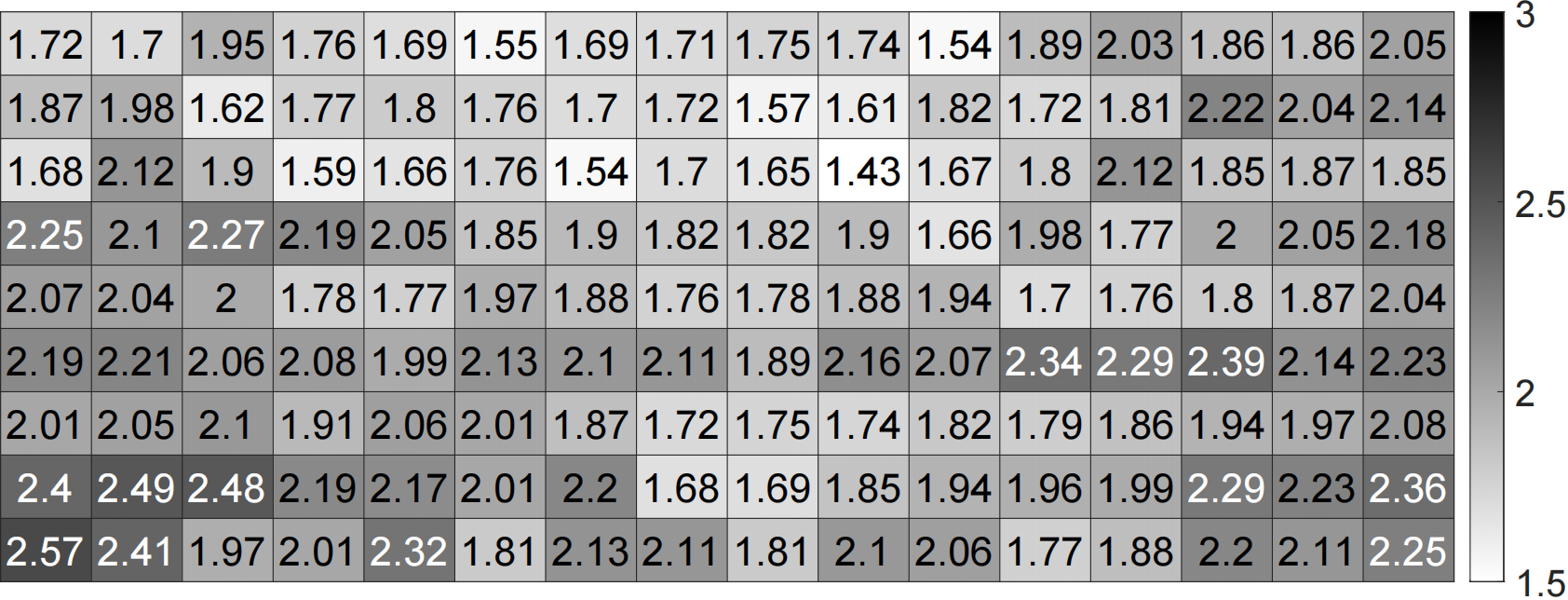}
         \vspace{-6mm}
        \caption{Screen-based gaze error distribution}
        \label{fig_screen}
    \end{minipage}
    \hfill
\end{figure}

\textbf{Gaze Error Range Model: }
We collected gaze data from 16 users to determine the gaze error range, as well as its screen-based distribution.
We use the same gaze tracking device, i.e., Tobii Pro Spark.
Each user collected 4,000 aligned gaze data samples via staring at visual anchors.
Figure \ref{fig_users} depicts the average gaze error of each user.
The overall average error among all users is 1.9455 cm.
With the gaze point estimated by Tobii Pro Spark and the screen-based coordinate of the visual anchors as ground truth, we analyse the gaze error range distribution.
We aggregate all users' gaze error and calculate its distribution on the 1920×1080 screen.
Figure \ref{fig_screen} demonstrates the screen-based error distribution.
We can see the gaze error tends to be higher near the borders, especially the left and right sides.
Additionally, higher error concentrates in the bottom area, which is close to the Tobii Pro Spark hardware.
We save this distribution map as the gaze error range model, and use it to dynamically adjust the estimated gaze error range based on the gaze location.
When the gaze vertically escapes the current line for over 2.5 seconds, we consider it an event of jump reading.

\subsection{Jump Reading Relocation} \label{sec_desgin_relocation}
Now that the jump reading has been initiated, the next step is to find the destination of the jump reading, i.e., to anticipate where the user is reading next.
Relocating the reading using only gaze information is almost impossible, since gaze error range largely exceeds line spacing.
Fortunately, we observe an opportunity hidden within the language context that can potentially aid jump reading relocation:
Punctuation marks serve as logical anchors in writing \cite{nunberg1990linguistics}. 
They help organize and clarify the meaning of sentences by indicating pauses, separating ideas, and showing relationships between different parts of the text. 
This makes users more likely to follow punctuation marks for efficient review or preview.
Even if a user does not specifically follow any punctuation mark as the starting point for review or preview, it is very likely that the user will soon encounter a punctuation mark on the way of resumed reading, which brings the situation back to the assumption that punctuation marks can be used as anchors.

\subsubsection{Relocation Candidate Identification}
As a result, we use punctuation marks as guidance for jump reading relocation.
After jump reading is detected, as depicted in Figure \ref{fig_jump}, we record the trajectory of the user's gaze, and search for punctuation marks that fall within the error range of the trajectory as candidates for potential relocation destinations.
As mentioned in Section \ref{sec_z_cut}, humans read from left to right, and looking from right to left for a long distance would highly possibly indicate line switching.
In other words, Z-cuts serve as effective indicators showing that the user has finished relocating and has resumed linear reading at a new starting point.
Therefore when a Z-cut is detected, we review the punctuation marks captured along the trajectory and begin determining the new starting point of linear reading.

\subsubsection{Candidate Election Algorithm}
If there is only one candidate captured with the error range of the trajectory, we use that punctuation mark as the new starting point to resume linear reading tracking.
However, if there are two or more candidates captured, it would be uncertain which one of them is the correct starting point.
To that end, we design a candidate election algorithm to find the best among them, using both gaze tracking results and contextual information as reference.

As punctuation marks serve as language anchors, each candidate has a sentence following it.
We first use the location and length of the candidate sentences to calculate their match ratio with the trajectory.
An intuitive solution to evaluate the match ratio would be calculating the area of overlapping between the sentence and the error range of the trajectory.
However, the previously proposed gaze error range model is merely a shape that lacks detailed information, such as how close the sentence is to the trajectory.
To fill this void with useful information, we further propose a finer gaze error vector distribution model, which also originates from the gaze data collected from the 16 users.

\begin{figure}[ht]
    \hfill
    \begin{minipage}[t]{0.66\linewidth}
        \centering
        \includegraphics[width=\textwidth]{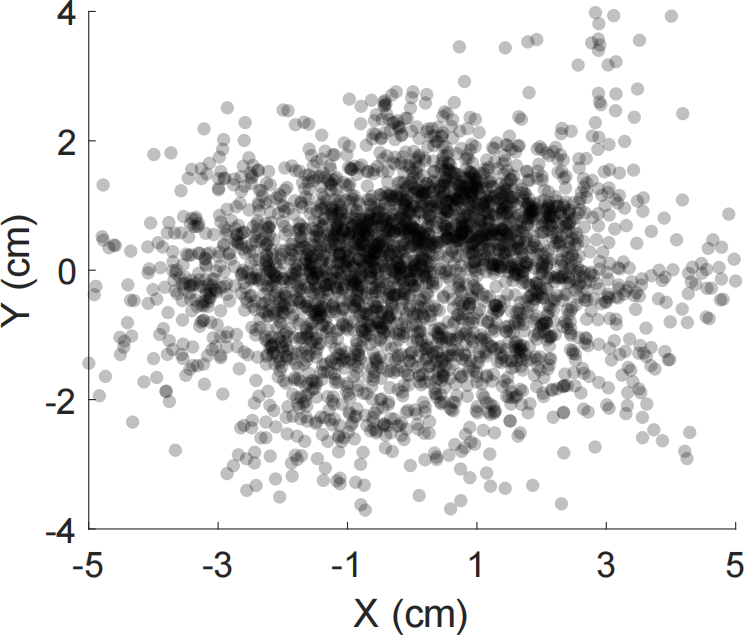}
         \vspace{-7mm}
        \caption{Error vector distribution}
        \label{fig_probab}
    \end{minipage}
    \hfill
    \hfill
\end{figure}

\textbf{Gaze Error Vector Distribution Model: }
To better model the probability of a candidate sentence being the actual sentence that the user is reading, we conduct real user study to analyse the distribution of the gaze error vectors, i.e., vectors from the gaze ground truth location to the estimated gaze location.
We aggregate the gaze error vector from all users' data, as depicted in Figure \ref{fig_probab}.
We can see the error vector distribution forms an oval cloud, where samples concentrate around the center.
Specifically, its horizontal and vertical standard deviations are 1.8471 cm and 1.2289 cm, respectively.
This distribution aggregates the gaze error vector samples regardless of the screen-based gaze location, as it would be too sparse if we also consider the screen-based location.
To ensure computational efficiency, we further randomly collect 500 samples from the cloud and save them as the gaze error vector distribution model.
To use the model for match ratio calculation, for any specific candidate sentence, and for every gaze location recorded within the jump reading trajectory, we attach the model onto it, and count what percent of the cloud samples falls on a candidate sentence.
We then accumulate the results of all gaze locations in the trajectory as the value of the match ratio for that candidate sentence.
This match ratio will then serve as a metric of evaluation for the candidates' election.

\subsubsection{Large Language Model Assisted Decision Making}

The match ratio algorithm alone could only leverage the spatial relationship between the jump reading trajectory and candidate sentences following punctuation marks.
This reminds us that we are handling reading materials, which hold rich contextual information that could also provide guidance for the candidate election.
The recent advancement of large language models has introduced new possibilities for aiding reading tracking tasks with their superior language perception capabilities.
Therefore, to further exploit the contextual information embedded in the text, we propose using LLM to aid the jump reading candidate election.
The inputs provided to the LLM include the text paragraph, the reading history before the detection of jump reading, and the candidate sentences.
The expectation for the LLM agent is that it judges which candidate sentence is most likely to be the new starting point, based on logical and linguistic relationships.

\textbf{Absence of Ground Truth: }
Typical methods of using LLMs are fine-tuning the LLMs with abundant domain-specific datasets to transform their perception capabilities.
However, in the domain of reading tracking, collecting a massive ground truth dataset is extremely hard, as there is no reliable way of labeling the reading progress.
A substitute method may be asking the user to use the cursor to indicate the reading progress.
However, such a method will inevitably divert the user's attention to actively staring at the cursor in order to acquire accurate ground truth, which would then impair the natural reading pattern.
Similar reason applies to other substitutes, such as vocally reading the text out loud.
As a result, the absence of ground truth data eliminates the possibility of fine-tuning the LLM to assist reading tracking.

\textbf{Prompt Engineering: }
The inherent language perception capabilities of LLMs are sufficient for understanding reading material and aid in electing candidates for jump reading, making fine-tuning less necessary. 
Therefore, we employ prompt engineering techniques to leverage the LLM's perception abilities. 
Prompt engineering capitalizes on the foundation model's pre-encoded knowledge, which can be activated by a specific prompt describing the task. 
To implement this, we augment all samples with a leading prompt, such as:

\vspace{5mm}
\textit{``The user was just reading: $<$Reading Material$>$, which option is most likely to be read next by the user?''}
\vspace{5mm}

By reformulating input samples as cloze-style phrases, we align them with the model's pre-training for text-based question answering. 
Before answering this question, the LLM receives the whole knowledge of reading material.
Additionally, we always choose the top three candidates with highest match ratio evaluation scores to feed the LLM, in order to reduce the complexity of the dialog.

In the final candidate election process, the LLM-chosen candidate receives an evaluation bonus. 
Typically, a high match ratio, evaluated by the gaze error vector distribution model, is around 0.3 out of 1.0. 
We assign a 0.1 bonus to the chosen candidate. 
This bonus, combined with trajectory-based matching evaluation, determines the highest-scoring candidate. 
The punctuation mark associated with this candidate then serves as the new starting point to resume linear reading tracking until the next jump reading event. 
By integrating prompt engineering with LLMs, we achieve a more efficient and accurate method for selecting jump reading candidates, leveraging the model's inherent capabilities without the need for extensive fine-tuning.

\subsection{Dynamic Calibration via Line-Gaze Alignment} \label{sec_desgin_calibration}
\begin{figure}[ht]
    \hfill
    \begin{minipage}[t]{0.75\linewidth}
        \centering
        \includegraphics[width=\textwidth]{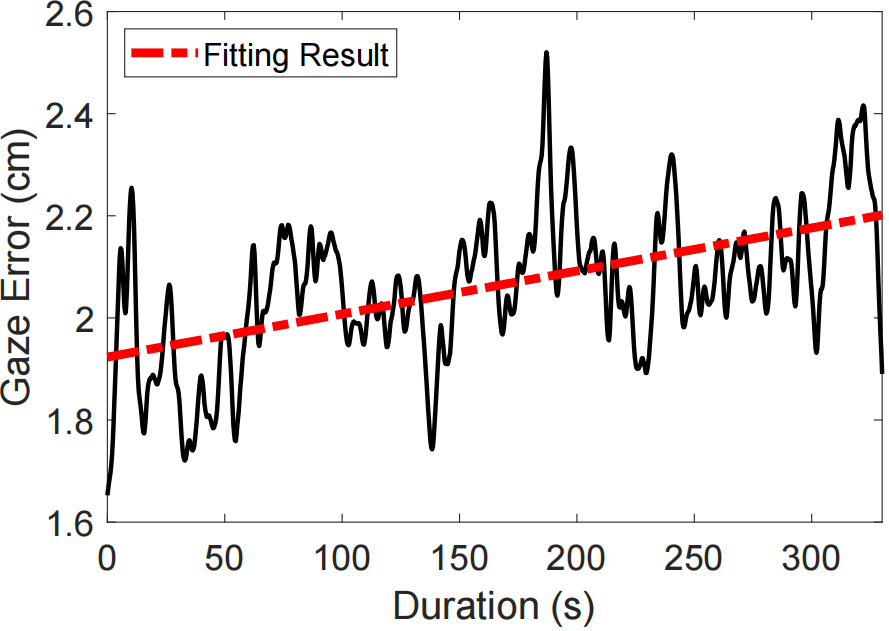}
         \vspace{-6mm}
        \caption{Calibration drifting}
        \label{fig_drift}
    \end{minipage}
    \hfill
    \hfill
\end{figure}

Reading is a relatively time-consuming task.
Based on the field test (Section \ref{sec_field_test}), the average time spent on reading a 200-word paragraph is 138.9 seconds.
During such a time span, we often find the users slightly and inadvertently drifting away from the position, where they conducted the calibration with the eye tracker.
This could cause the gaze tracking calibration results to gradually degrade or fail, resulting in a slowly increasing gaze tracking error.
To demonstrate the effect of drifting, we aggregate the gaze data traces of all 16 users along one time axis to reflect the overall temporal trend of gaze tracking error.
As depicted in Figure \ref{fig_drift}, we can see that the gaze tracking error gradually increases over time.
According to the linear fitting results, the gaze error could grow from 1.9244 cm to 2.2015 cm within 330 seconds.

This drifting issue necessitates additional re-calibration procedures to frequently correct the gaze tracking results.
However, classical gaze tracking calibration methods require active attention from the users, forcing them to stop the current reading activity and engage in a relatively tedious calibration process.
In pursuit of maintaining gaze tracking accuracy as well as saving users from active calibration requirements, we propose a dynamic calibration method that makes use of the calibration opportunity exclusively embedded in the domain of reading tracking.

During linear reading, the user can be assumed to be actively looking at the current line, while the alignment between the gaze and the line provides a valuable dynamic calibration opportunity.
Such alignment opportunities happen repeatedly during reading and do not require specific user participation.
Since the lines are horizontally oriented and vertically arrayed, they provide rich resources for vertical calibration.
Therefore, we mainly focus on calibrating the Y-axis (vertical) of gaze tracking results.
On the other hand, the horizontal gaze error does not cause great issues to the lines that are much wider.
We record the Y-axis of the raw gaze results during linear reading of each line.
When a line is finished, we calculate the average of the Y-axis raw gaze results, $Y_g$.
With the line's vertical location $Y_l$, they form a gaze-line calibration pair $<Y_g, Y_l>$.
With multiple gaze-line pairs sampled via reading multiple lines, we perform linear regression to fit the gaze results to the lines:
\begin{equation}
    [k, b] = \arg \min_{k, b} \Sigma(Y_l - (k \cdot Y_g + b ))^2
\end{equation}
We then apply the regression results $<k, b>$ to scale and de-bias the raw gaze tracking results $[X, Y]$ into calibrated gaze results $[X', Y']$:
\begin{equation}
    [X', Y'] \leftarrow [X, k \cdot Y + b]
\end{equation}
This dynamic calibration design can constantly calibrate the gaze tracking results and suppress its drift error, as long as the user is conducting linear reading.
Even in the event of jump reading, after the relocation and linear reading resumes, the dynamic calibration may also proceed.

    \section{Implementation}
\begin{figure}[ht]
    \hfill
    \begin{minipage}[t]{0.98\linewidth}
        \centering
        \includegraphics[width=\textwidth]{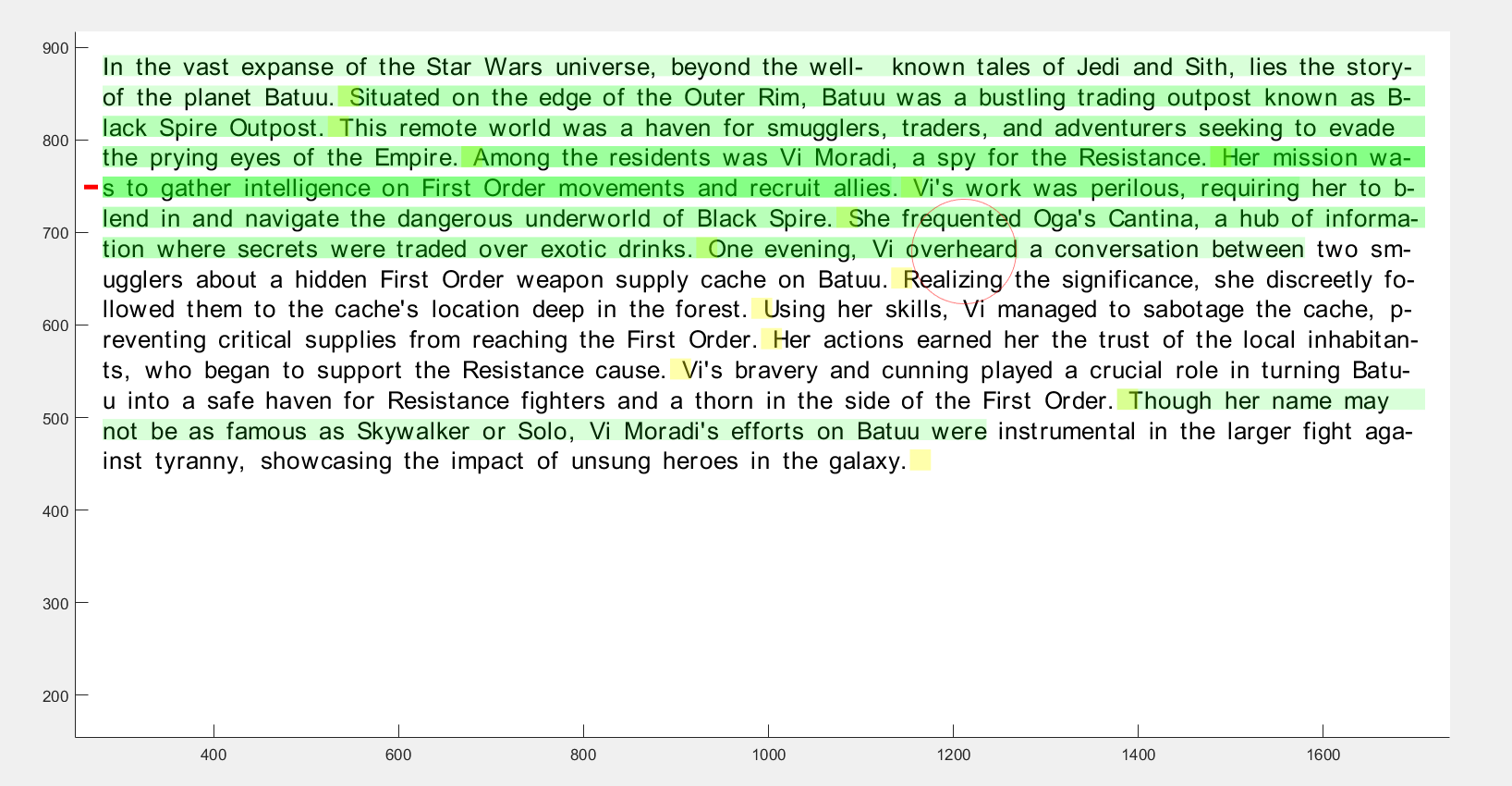}
         \vspace{-6mm}
        \caption{System demonstration}
        \label{fig_implem_demo}
    \end{minipage}
    \vspace{-3mm}
    \hfill
    \hfill
\end{figure}

We implemented \ours using Matlab, on a laptop with a 15.6" screen of 1920x1080 resolution.
The laptop is equipped with Intel(R) Core(TM) i7-8750 CPU @ 2.20GHz.
As demonstrated in Figure \ref{fig_implem_demo}, we render the text paragraph within a blank plot window, and thus acquire in first place the coordinate of all words, letters, and punctuation marks.
We set the default line spacing as 4.5mm.

\subsection{Gaze Tracking}
We use the state-of-the-art eye gaze tracker, Tobii Pro Spark \cite{tobii}, as gaze tracking input.
It uses infrared emitters and sensors to capture featured eye reflection to track gaze.
With the infrared feature, it can work in all lighting conditions, including dark environments.
We follow the standard calibration procedure of the Tobii Pro Eye Tracker Manager software to calibrate the Tobii Pro Spark for every user before the test.

\subsection{Text Highlighting}
We highlight the words that have been read by the user according to the reading tracking results.
The highlighting is implemented via a light green area under the text with 80\% transparency, serving as visual guidance without greatly distracting the users.
Reading the same words multiple times will enhance the existing highlighting.
Additionally, upon system initialization, we highlight all punctuation marks within the text material using yellow shades, specifically periods, exclamation marks, and question marks.

\subsection{Large Language Model}

To enhance the capabilities of \ours, we integrate Matlab and Python to access the OpenAI GPT-API.
This integration allows us to leverage the advanced language processing capabilities of state-of-the-art language models.

Specifically, for the large language model, we use GPT-4o mini. GPT-4o mini is a variant of the GPT-4 model, designed to provide high performance while being more efficient in terms of computational resources. 

\subsection{Forced Relocation via Double Clicking} \label{imple_double_click}
In the case that \ours is not tracking and highlighting accurately, the users are provided with a function of forced relocation by double clicking.
When they double click on a word, the tracking will strictly relocate onto that word and resume tracking.
The system will return a beep sound as confirmation to the user if the double clicking lands on a valid word, instead of another invalid blank area. 
The jump reading detection will reset upon the double clicking as we assume the user will look at that word when performing double clicking.
The frequency of this function being used also reflects the reliability of \ours.
Lower frequency indicates lower demand of user participation, as well as potentially higher satisfaction.
    \section{Controlled Experiment}
In this section we conduct controlled experiments to statistically evaluate the reading tracking accuracy.

\subsection{Experiment Setting}
We collect reading tracking ground truth via asking the user to look at the cursor, and use it to indicate the progress of reading.
However, this ground truth data collection method inevitably impacts the natural reading pattern, making certain evaluations relatively unrealistic.
In another word, we could only simulate reading when using cursor to collect ground truth.
We rely on the field test for more real-case evaluations.

\begin{figure}[ht]
    \hfill
    \begin{minipage}[t]{0.6\linewidth}
        \centering
        \includegraphics[width=\textwidth]{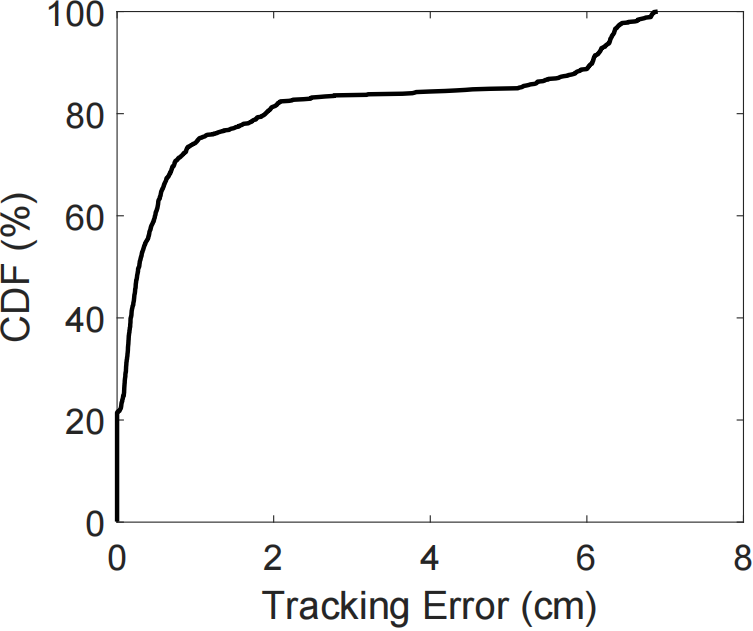}
         \vspace{-6mm}
        \caption{Linear reading tracking}
        \label{fig_expm_linear}
    \end{minipage}
    \vspace{-3mm}
    \hfill
    \hfill
\end{figure}

\subsection{Linear Reading Tracking Accuracy}
We first evaluate the reading tracking accuracy during linear reading.
We set the evaluation metric as the distance between the current word being read according to the reading tracking result, and the ground truth reading progress labeled by the cursor.
Specifically, if the cursor falls exactly on the word that is anticipated by the tracking algorithm as the current word being read, we mark the tracking error as zero.
As shown in Figure \ref{fig_expm_linear}, the reading tracking error concentrates on both low and high levels.
It indicates that \ours is able to firmly track the words in the same line, while the high error is probably due to line switching, during which the gaze needs to move a long distance from right to left.
The overall average linear reading tracking error is 1.3161 cm, far lower than the 2 cm-level raw gaze tracking error, demonstrating the advantage of adopting linear reading assumption.

\begin{table}[t] 
    \centering
    \caption{Jump reading tracking accuracy} \label{table_jump}
    \scalebox{1.2}{
        \begin{tabular}{|c|c|c|c|c|}
            \hline
            Candidates & 1   & 2   & $\geq$3  & Overall  \\ \hline
            Accuracy  &  94.44\%   &  84.62\%   &  57.14\%  &  84.21\%  \\ \hline
        \end{tabular}
    }
    \vspace{-4mm}
\end{table}

\subsection{Jump Reading Tracking Accuracy}
We evaluate the accuracy of jump reading tracking via checking whether the jump reading relocation arrives at the correct line.
We also monitor the amount of candidates captured along the jump reading trajectory to study its impact on the relocation accuracy.
As shown in Table \ref{table_jump}, the overall accuracy of jump reading relocation is 84.21\%, and the accuracy is 94.44\% when there is only one candidate.
The gap between 94.44\% and 100\% originates from the inherent gaze tracking error that could cause the trajectory to miss the correct candidate.
When there are two candidates contending for the election, the accuracy is 84.62\%.
Usually the match ratio of the correct candidate would largely exceed the other candidate, making the election relatively robust.
However, when there are three or more candidates captured along the trajectory, the accuracy drops to 57.14\%.
Multiple candidates being considered usually implies that the destination area is crowded with candidates, whose locations may fit the gaze error vector model with similar match ratio, making the election difficult to distinguish the correct candidate.

\begin{figure}[ht]
    \hfill
    \begin{minipage}[t]{0.85\linewidth}
        \centering
        \includegraphics[width=\textwidth]{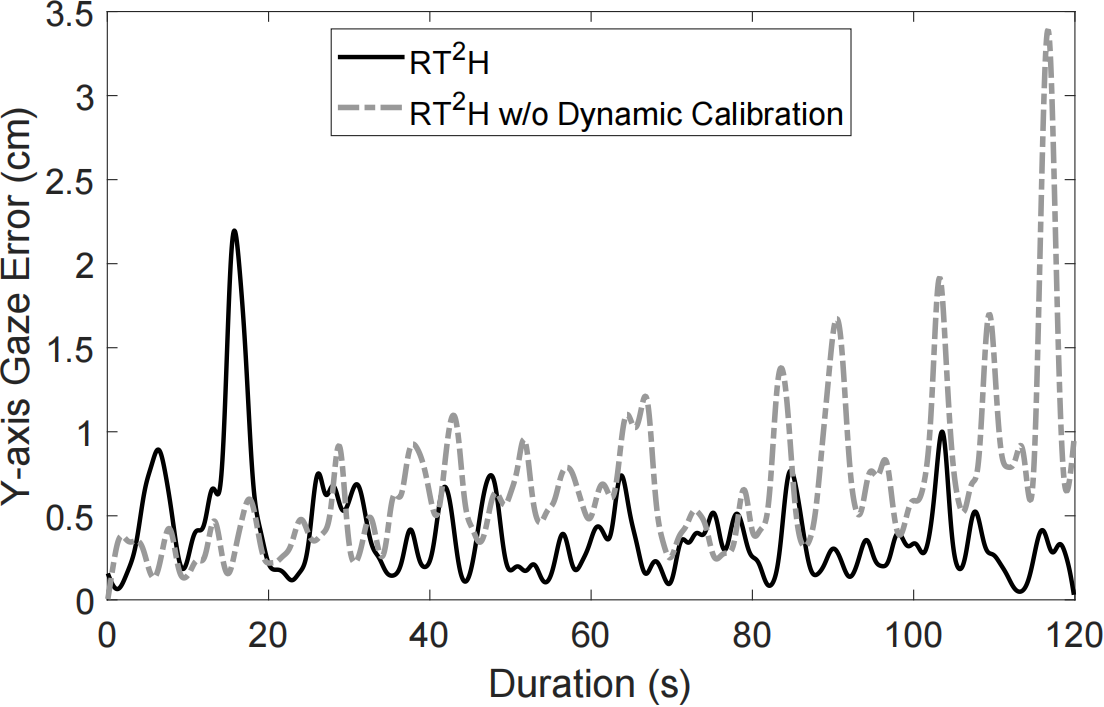}
         \vspace{-6mm}
        \caption{Dynamic calibration ablation study}
        \label{fig_expm_dyna_cali}
    \end{minipage}
    \vspace{-3mm}
    \hfill
    \hfill
\end{figure}

\subsection{Ablation Study on Dynamic Calibration}
To examine the drift-suppressing effect of the dynamic calibration component, we compare \ours against \ours without the dynamic calibration mechanism ($\ours \ w/o \ DC$).
We test the two baseline systems on the same reading material for 20 times, and aggregate all data traces onto one time axis to reflect the overall error trend of the two baseline systems.
Specifically, we focus on the Y-axis (vertical) gaze error.
As shown in Figure \ref{fig_expm_dyna_cali}, despite an error peak at the beginning, \ours demonstrates higher stability when compared to the $\ours \ w/o \ DC$.
Without the drift suppression enabled by the dynamic calibration design, it tends to destabilize after 60 seconds of active usage.
Statistically, the average Y-axis error over 120 seconds is 0.3904 cm, and 0.6609 cm, for \ours and $\ours \ w/o \ DC$, respectively.

\subsection{Computation and Communication Latency}
Based on real experiment, we observe that every iteration in linear reading tracking takes an average of 0.000127 second to finish, which potentially supports up to 7500+ Hz inference.
In the actual implementation, we added a forced pause of 0.05 second for linear reading tracking to cap its iteration frequency.
As for jump reading relocation, the candidate election procedure takes an average of 0.7893 second to complete, with 0.6382 second of it consumed by LLM API communication.
Such latency does not greatly impact users' experience.

    \section{Field Test} \label{sec_field_test}
We propose the real-time highlighting function in order to help users track their reading progress and reduce the chance of switching to wrong lines after finishing reading one.
However, the real-time highlighting is a dual-way interaction.
Both the system and the user receives feedback from another in real-time. 
Such attention-sensitive interaction requires users to fully immerse themselves in the system, making simulated experiments insufficient.
Therefore, we put \ours into the real field test, to evaluate the combined effect of reading tracking and real-time highlighting.

We invited 18 volunteers to test \ours.
Specifically, the volunteers are asked to read and understand certain text paragraph.
To ensure engagement, each reading material includes a question with an answer embedded in the paragraph. 
The questions are straightforward, and answers are apparent in the text. 
However, the volunteers need to search among the lines for the answer, necessitating the action of review and preview.
We recorded the time taken to find the correct answer, informing volunteers if their initial response was incorrect and continuing the timing until they provided the correct answer.



\subsection{Baselines}
\begin{itemize}
    \item \ours: The full version of our reading tracking and real-time highlighting system, whose design components are introduced in this paper.

    \item $RTH$: The simple version without reading tracking components.
    It directly uses the gaze tracking results to highlight the words that the gaze falls onto. 
    By comparing $RTH$ and \ours, we demonstrate the effectiveness of our design components.
    
    \item Blank: The blank version, where the system only renders the paragraph with the provided text material, using the same rendering format as other baselines, e.g., font size, line spacing, paragraph width, etc..
\end{itemize}


We recognize that the order in which the versions are tested could impact results. 
For example, volunteers may feel unfamiliar with the hardware, and could potentially spend more time in the first test.
This would make it unfair if any version is always tested first.
Therefore, we shuffle the order of the versions being tested, resulting in six different testing orders: `017', `071', `107', `170', `701', `710', with number `0', `1', and `7' representing the version number.
This is also the reason we recruited $3\times6=18$ volunteers.
Each testing order was applied on three different volunteers, ensuring fair comparison among the three versions.

\subsection{Reading Materials}
Since we assigned questions for the volunteers to answer, each reading material would expire for that volunteer after being read, and cannot be used twice for the same user.
Additionally, there are three versions of the system to be tested.
Therefore, we arranged three different reading materials for the reading test.
The three reading materials contain 272, 257, and 249 words, respectively.




\begin{table}[t]
\centering
\caption{Comparisons among baselines} \label{tbl:field-test}
\scalebox{1.2}{
\begin{tabular}{|c|c|c|c|}
\hline
Algorithm       & \ours & RTH & Blank \\ \hline
Required Time   & 98.2 s    &  115.7 s   &  113.5 s    \\ \hline
User Experience &  3.8$\pm$1.0   & 2.7$\pm$1.4    &  3.3$\pm$1.4   \\ \hline
\end{tabular}
}
\vspace{-3mm}
\end{table}

\subsection{Reading Efficiency}
We evaluate the reading efficiency with different versions of the system via recording the time spent by the volunteers to come up with the right answer.
As illustrated in Table~\ref{tbl:field-test}, our algorithm, \ours, demonstrated satisfying performance compared to the baselines.
\ours required an average time of 98.2 seconds, compared to 113.5 seconds for blank version and 115.7 seconds for $RTH$.
This verifies that highlighting text could potentially enhance reading efficiency.


\subsection{User Experience on Different Baselines}
After volunteers finished testing all three versions, we asked their experience (ranges from 1 to 5) of using the systems via a questionnaire.
$RTH$ achieved a score of 2.7, which is even lower than the score of 3.3 of the blank version.
This indicates that directly applying the gaze tracking results to highlight text could produce negative effects.
Meanwhile, \ours achieved the highest overall user experience score of 3.8, showcasing the effectiveness of the algorithmic designs of \ours.

\subsection{User Evaluation on System Features}
We also collected feedback on several important features of \ours, including the system's helpfulness, whether it causes distraction, accuracy of jump reading tracking, and overall interest to see it on the market.
Overall, \ours's strong performance across these features underscores its robustness and efficiency.
These findings suggest \ours’s potential for broader application in real-world scenarios.

\subsubsection{Whether Highlighting Is Distraction}
To address a common concern on whether the highlighted area would distract the user, we add in the questionnaire a specific question on whether it distracts them.
We receive an evaluation feedback of $3.8\pm1.1$, indicating the highlighting does not cause great distraction, possibly due the 80\% transparency.

\subsubsection{Jump Reading Tracking}
We specifically notify the volunteers that \ours supports jump reading.
After the test, we investigate their satisfaction on the jump reading tracking accuracy, for which \ours receives a score of 3.6, indicating relatively reliable jump reading tracking support.

\begin{table}[t]
\centering
\caption{Statistics of questionnaire} \label{tbl:field-test-2}
\scalebox{1.3}{
\begin{tabular}{|c|c|}
\hline
Feature                & User Evaluation     \\ \hline
Reading Assistance     &  4.0$\pm$0.5    \\ \hline
Non-distracting        & 3.8$\pm$1.1  \\ \hline
Jump-reading Tracking    &  3.6$\pm$0.8    \\  \hline
Market Anticipation     &  3.4$\pm$1.5  \\  \hline
\end{tabular}
}
\vspace{-3mm}
\end{table}

\subsection{Force Relocation Frequency}
As introduced in Section \ref{imple_double_click}, the volunteers are provided with a function of force relocation by double clicking.
Lower frequency of force relocation being performed would indicate higher reliability of our system.
We make a clear introduction on this function to every volunteer before the testing of \ours.
During the test, we count the amount of valid double clicks by listening to the confirmation beep.
According to the test records, the average number of double clicking is 0.28, indicating relatively high robustness of \ours.

    \section{Related Work}

\subsection{Eye gaze tracking.}

Eye gaze tracking has garnered significant research interest due to its diverse applications in human-computer interaction \cite{duchowski2007eye, majaranta2014eye, salvucci2000identifying, borji2014defending, holmqvist2011eye}, psychology \cite{liversedge2000saccadic, hessels2019eye, holmqvist2017eye}, and medicine \cite{borji2014defending, wu2008detecting, pfeiffer2013gaze, vrij2000detecting}. Recent studies have explored innovative applications of this technology in interactive systems and wearable devices. For instance, Khamis et al. introduced GazeTouchPIN, a multimodal authentication system integrating gaze input and touch interaction to protect sensitive data on mobile devices \cite{huang2018gazetouchpin}. Jansen et al. presented EyeScout, a system for active eye tracking enabling position-independent gaze interaction with large public displays \cite{jansen2014eyescout}.

Geometric methods are a fundamental approach in gaze tracking, leveraging the geometric relationships between the eye and a camera or sensor to estimate gaze direction \cite{duchowski2007eye, holmqvist2011eye,  chamberlain2007eye}. These methods rely on precise calibration and mathematical models to accurately infer gaze direction. Early techniques, such as the corneal reflection method \cite{holmqvist2011eye}, used the reflection of light on the cornea and pupil to estimate gaze direction \cite{duchowski2007eye, chamberlain2007eye}. By analyzing the positions of corneal reflections relative to known eye or environmental features, researchers could determine gaze direction with reasonable accuracy. However, these methods required controlled lighting and were sensitive to head movements.

Advancements in geometric gaze tracking have led to the development of non-invasive, camera-based systems capable of real-time gaze direction estimation. Model-based tracking techniques, using 3D models of the eye and surrounding features, have improved accuracy and robustness \cite{majaranta2014eye}. AS-Gaze, for example, uses an iris model to estimate the gaze ray, allowing gaze tracking on various surfaces such as mobile phone screens, computer displays, or whiteboards \cite{ASgaze}. However, AS-Gaze faces limitations with free movement, depth information acquisition, and real-time performance, as it uses a single camera and requires significant computational time to match the iris boundary, resulting in an inference rate of only 11Hz, which is insufficient for real-time applications.

Parallel to these geometric methods, deep learning approaches \cite{dxz_drl_1, dxz_drl_2, dxz_drl_3} have also been explored in gaze tracking \cite{Balim_2023_CVPR, DVgaze, park2020eve}. A notable example is DV-Gaze, which uses multiple cameras to estimate gaze points and manage free movement by perceiving depth. DV-Gaze employs ResNet \cite{resnet} for face image processing and Transformers \cite{transformer} to encode camera calibration results, thereby handling the geometric relationships between cameras. Despite its innovative approach, DV-Gaze's computational intensity and reliance on complex face images pose challenges for practical deployment, leading to heavy training overhead and reduced feasibility for real-world applications.

\subsection{Reading tracking and LLMs}
Eye-gaze tracking for reading progression has seen innovative methods aimed at mitigating noisy data and the limitations of commercial eye-tracking devices. Bottos and Balasingam \cite{bottos2019novel} introduced a Slip-Kalman Filter for tracking reading progression, significantly improving line detection accuracy and noise reduction over standard Kalman filters. However, this method assumes sequential reading without repetition, which may not reflect natural reading behaviors. Alternative methods like hidden Markov models \cite{kayaalp2022hidden,wei2022provably,chen2023learning} for classifying eye-gaze fixation points \cite{bottos2019approach} improve line detection accuracy but require initial parameter estimation and a consistent text layout, limiting their flexibility. Least squares batch estimation techniques have been used to filter noisy data \cite{bottos2019tracking}, enhancing reading smoothness but facing limitations in initial line detection accuracy and managing repeated readings of the same line. Combining least squares batch estimation with hidden Markov models offers smoother progression but necessitates predefined line numbers and initial training with synthetic data \cite{bottos2020tracking}.

A significant challenge in the field is addressing jumping reading behaviors, where readers skip forward or backward within a text. This non-linear reading pattern, common in real-world scenarios, challenges the effectiveness of sequentially oriented methods like the Slip-Kalman Filter. This highlights the need for more flexible and adaptive eye-gaze tracking methods. Therefore, we propose integrating LLMs into our design, leveraging their adaptability in understanding and generating human-like text \cite{xu2023exploring,qiao2024latency,chen2024communication,du2024age,vierling2024input,yang2024drhouse}. While fine-tuning LLMs involves extensive data collection \cite{huang2023towards}, we employ prompt engineering to guide the LLM in generating desired outputs without extensive retraining \cite{siledar2024one,hu2024improving,kan2023prompt,xu2024prompting}. This approach enhances the flexibility and adaptability of eye-gaze tracking systems, particularly in managing non-linear reading behaviors.

Prompt engineering has been successfully applied in various domains to enhance the performance of LLMs. For instance, in clinical named entity recognition, detailed instructions and contextual specifications significantly improve the extraction of critical medical information from clinical notes and safety reports \cite{hu2024improving}. Similarly, LLMs have been used to predict chemical properties and optimize experimental conditions \cite{kan2023prompt}. Xu et al. \cite{xu2024prompting} proposed an advanced framework that leverages prompt engineering to boost the performance of LLMs in recommender systems. This framework employs carefully designed prompts to enhance tasks such as click-through rate prediction. These approaches demonstrate the dual benefits of reducing training efforts and leveraging pre-trained language perception capabilities.

Addressing practical challenges in prompt design, Pereira et al. \cite{pereira2023why} introduce a no-code chatbot design tool that enables non-AI experts to create and evaluate prompts through iterative design and testing. Additionally, Zhang et al. \cite{zhang2024vpgtrans} present a method for transferring visual prompt generators across different LLMs to reduce computational costs, using a two-stage transfer framework. For aligning graph information with LLMs, Liu et al. \cite{liu2024can} propose the use of soft prompts combined with graph neural networks to address the modality discrepancy between graph and text data.

    \section{Conclusion}

In this paper, we present \ours, a reading tracking and real-time highlighting system designed to support both linear and jump reading. 
\ours adopts experimental gaze study on 16 users to support detection and relocation of jump reading, and further leverages large language models to assist jump reading tracking using its contextual perception capability.
Controlled experiments showed that \ours achieves an accuracy of 84\% in jump reading relocation.
Field tests with 18 volunteers confirmed the system's practicality, demonstrating a 13.5\% reduction in reading time compared to baselines. 
The positive user feedback further validates the effectiveness of \ours in real-world applications.


\end{document}